\documentclass[preprint,showpacs,preprintnumbers,amsmath,amssymb]{revtex4}

\usepackage{graphicx}

\usepackage{natbib}
\usepackage{multirow}
\usepackage{latexsym,amssymb}
\usepackage{amsmath}

\newcommand{\Rpp}{\vec{R}_{\perp}}
\newcommand{\kpp}{\vec{k}_{\perp}}
\newcommand{\kpr}{\vec{k}_{\parallel}}
\newcommand{\rpr}{\vec{r}_{\parallel}}
\newcommand{\Pel}{\vec{P}_{el}}
\newcommand{\ver}{\vec{r}}
\newcommand{\veR}{\vec{R}}
\newcommand{\vek}{\vec{k}}

\begin{document}

\title [Polarization structure]{The structure of electronic
polarization and its strain dependence}
\author{Yanpeng Yao} \author{Huaxiang Fu}
\affiliation{Department of
Physics, University of Arkansas, Fayetteville, AR 72701, USA}
\date{\today}

\begin{abstract}
The $\phi(\kpp)\sim \kpp$ relation is called polarization structure.
By density functional calculations, we study the polarization structure
in ferroelectric perovskite PbTiO$_3$, revealing (1) the $\kpp$ point that
contributes most to the electronic polarization, (2) the magnitude of
bandwidth, and (3) subtle curvature of polarization dispersion.  We also
investigate how polarization structure in PbTiO$_3$ is modified by
compressive inplane strains. The bandwidth of polarization dispersion in
PbTiO$_3$ is shown to exhibit an unusual decline, though the total
polarization is enhanced. As another outcome of this study, we
formulate an analytical scheme for the purpose of identifying what
determine the polarization structure at arbitrary $\kpp$ points
by means of Wannier functions. We find that $\phi(\kpp)$ is determined
by two competing factors: one is the overlaps between neighboring Wannier
functions within the plane {\it perpendicular} to the polarization direction,
and the other is the localization length {\it parallel} to the polarization direction.
Inplane strain increases the former while decreases the latter, causing
interesting non-monotonous effects on polarization structure.
Finally, polarization dispersion in another paradigm
ferroelectric BaTiO$_3$ is discussed and compared with that of PbTiO$_3$.

\pacs{77.22.Ej, 77.80.-e}
\end{abstract}

\maketitle

\section{Introduction}

Electric polarization is a key quantity for computing and
understanding technologically-relevant effective charges, dielectric
and piezoelectric responses that are the derivatives of polarization
with respect to atomic displacement, electric field, and strain,
respectively.\cite{Lines} Polarization also plays an important role
in the methodology development of the theory dealing with finite
electric fields in infinite solids, by minimization of the free
energy $F=U-\vec{E}\cdot{\vec{P}}$~\cite{Souza,Umari,Fu_E,Dieguez}.
Total electric polarization consists of electronic contribution
($\Pel$) and ionic component ($\vec{P}_{ion}$). Computing the latter
component is straightforward using point charges, while calculating
the electronic polarization is not. Today $\Pel$ is calculated using
the sophisticated modern theory of
polarization\cite{King-Smith,Resta}. According to the theory, $\Pel$
corresponds to a geometrical phase of the valence electron states,
\begin{equation}
   \Pel=\frac{2e}{(2\pi)^3}\int d\kpp\phi(\kpp) \ \ ,
\label{EPel}
\end{equation}
where
\begin{equation}
 \phi(\kpp)=i\sum_{n=1}^M\int_0^{G_{\parallel}}d\kpr\langle
u_{n\vek}|\frac{\partial}{\partial k_\|}|u_{n\vek}\rangle
\label{EPhi_1}
\end{equation}
is the Berry phase of occupied Bloch wave functions $u_{n{\vek}}$.
Subscripts $\parallel$ and $\perp$ mean parallel and perpendicular
to the polarization direction, respectively. Practically,
to carry out the $\Pel$ calculations,
the integral in Eq.(\ref{EPel}) is replaced by a weighted summation of
the phases at sampled discrete $\vek$-points
(Monkhorst-Pack scheme\cite{Monkhorst}, for example) in the 2D $\kpp$ plane,
namely, $\Pel=\sum_{\kpp}\omega (\kpp)\phi(\kpp)$ with weight $\sum_{\kpp}\omega (\kpp)=1$.
The polarization at individual $\kpp$, $\phi(\kpp)$,
is calculated as the phase of the determinant formed
by valence states at two neighboring $\vek$s on the $\kpr$ string
as \cite{King-Smith,Resta}
\begin{equation}
\label{EPhi_2}
    \phi(\kpp)={\rm Im}\{\ln\prod_{j=0}^{J-1}\,
    {\rm det}(\langle u_{k_j,m}| u_{k_{j+1},n}\rangle)\}  \ \ .
\end{equation}
Defined as such, the total
polarization $\vec{P}=\vec{P}_{ion}+\Pel$ could be uniquely
determined and gauge independent up to a modula constant. In
Eq.(\ref{EPel}) one sees that, it is the $\phi(\kpp)$ phases
at different $\kpp$ that determine the electronic
polarization. The purpose of this work is to study the properties
of $\phi(\kpp)$.

The physical significance of the $\phi(\kpp)$ quantity can be
understood by analogy. It is well known that band structure, which
describes the relation between single-particle orbital energy and
electron wave vector $\vek$, is very useful for understanding
electronic, photo-excitation, and photoemission properties in
solids\cite{Yu}. The $\phi(\kpp)\sim \kpp$ relation may be similarly
termed as ``polarization structure'', or ``polarization-phase
structure''. Electron states in band structure can be changed by
photo-excitation or emission.  The $\kpp$-point polarization phase
can be altered by electric fields, which act as a possible
excitation source for electrical polarization. Note that electrical
fields do not alter the electron wave vector ($\kpp$) perpendicular
to the direction of the field, and thus $\kpp$ remains a conserved
quantity. The field-induced variation of $\phi(\kpp)$ in fact
manifests the $\kpp$-dependent polarization current. As a result,
the relevance of polarization structure to electronic polarization
is like the band structure to electronic properties.

Furthermore, understanding the $\phi(\kpp)$ quantity is of useful
value from both fundamental and computational points of view.
Fundamentally, this quantity is determined by the Bloch wave
functions, not in the ordinary sense of spatial distribution, but
through the interesting aspects of the Berry's phase of occupied
manifold of electron states. Studying how $\phi(\kpp)$ depends on
$\kpp$ may yield better understanding of electron states, as well as
the rather intriguing connection between these states and their
contributions to polarization in insulator solids.  Computationally,
we first recognize that the $\phi(\kpp)$ phase computed from
Eq.(\ref{EPhi_2}) always produces a value within the principal range
$[0,2\pi]$. In reality, depending on the dispersion of $\phi(\kpp)$
as a function of $\kpp$, it is possible that the phases for
different $\vek$ points fall in different branches. In other words, the
true $\phi(\kpp)$ values may fall in the principal range for some
$\kpp$ points (Let us denote this set of $\kpp$ points as
$\kpp^{(I)}$), while falling out of the principal range for other
$\kpp$ (to be denoted as $\kpp^{(II)})$.  We find numerically that
this indeed happens for real materials particularly when
polarization is large; a specific example is given in section II.
When this occurs, one must {\it not} artificially shift the phases
of the $\kpp^{(II)}$ into the principal range, as computers do
according to Eq.(\ref{EPhi_2}). Though this shift makes no
difference to the polarization phase of individual $\kpp$ points, it
will alter the total $\Pel$ polarization, yielding spurious
magnitude of polarization. Only when the phase of {\it every} $\kpp$
is shifted by a constant $2\pi$ will the total $\Pel$ polarization
remain equivalent. To find out which $\kpp$ may generate a phase not
in the principal range, one in principle should compute the whole
dispersion structure of polarization and then map out the
$\phi(\kpp)$ for all $\kpp$ points based on the assumption that the
$\phi(\kpp)$ phase is a continuous function of wave vector $\kpp$,
which makes it important to study the properties of the $\phi(\kpp)$
phase as a function of $\kpp$.

Despite the relevance, the dispersion structure of polarization is
nevertheless not completely understood. More specifically, (1) little is
known about what determine the $\phi(\kpp)$ phase at individual
$\kpp$. In Eq.(\ref{EPhi_2}), $\phi(\kpp)$ is determined by the wave
functions of a string of $\kpr$ points, not just a single $\vek$. As
a result, the answer to the question is highly non-trivial. (2) For
a given ferroelectric substance (say, the prototypical PbTiO$_3$),
it is not clear which $\kpp$ exhibits the largest polarization
contribution. Does the $\Gamma $ point always contribute most or
least? (3) We do not know if the Berry's phases at different $\kpp$s
share a similar value or are very different from each other, that
is, a problem concerning the dispersion width of the polarization
structure. Slightly more intriguing, one may wonder along which
direction the $\phi\sim \kpp$ curve shows the largest dispersion?
(4) Even for two commonly studied ferroelectrics, BaTiO$_3$ and
PbTiO$_3$, we do not know how different or similar their
polarization structures are.

Recently, there is another active field in the study of
polarization, which concerns the use of inplane strain to tune the
ferroelectric polarization \cite{Choi,Haeni,Ederer,Lee}. This
tunability stems from the fundamental interest in the
strain-polarization coupling. Imposed under inplane strain
ferroelectrics subject to modifications of chemical bonds and/or
charge transfer, thereby the interaction between atoms is altered.
It has been known that a compressive inplane strain tends to enhance
the total polarization. But the amplitude of enhancement was found
to be highly material dependent.\cite{Ederer,Lee} Considering the
importance of the strain effects, one might want to know how the
$\phi(\kpp)$ phase from each $\kpp$ can be influenced by strain.
Strain effects on the polarization dispersion remain largely
unknown, however. It would be of interest to examine how the strain
may tune and modify the dispersion of polarization structure.
Specific questions on this aspect are: in what manner would the
inplane strain change the relative contributions and curvatures at
different $\kpp$, and how the band width of the dispersion curve is
to be altered.

With these questions in mind, we here study the dispersion structure
of the polarization in ferroelectric perovskites, as well as its
dependence on inplane strains. Two complementary approaches
(first-principles density functional calculations and analytical
formulations) are used. By means of analytical formulation, we aim
at a better understanding on what specific quantities and/or
interactions determine the polarization at individual $\kpp$ point.
Our calculations reveal some useful knowledge on the
polarization structure in perovskite ferroelectrics. For example,
the largest $\phi(\kpp)$ contribution is shown not to come from the
zone center, but from the zone boundary. We also find that the
polarization curve in PbTiO$_3$ is notably flat along the
$\Gamma-X_1$ direction, and exhibits, however, a strong dispersion
along the $\Gamma-X_2$ axis. Our theoretical analysis further
reveals that the flat dispersion along the $\Gamma-X_1$ direction is
caused by a small amount of participation from the nearest-neighbor
interaction between the Wannier functions. Finally, the present
study also demonstrates some rather interesting differences in
PbTiO$_3$ and BaTiO$_3$, in terms of the polarization structures as
well as their strain dependences.

\section{The polarization structure of lead titanate}

We first present the density functional calculations on the
polarization structure in PbTiO$_3$. In its ferroelectric
phase PbTiO$_3$ is tetragonal ($|\textbf{a}_1|=|\textbf{a}_2|=a, |\textbf{a}_3|=c$)
and possesses a large spontaneous polarization.  The
polarization is along the $c$-axis direction, perpendicular to the
$\kpp$ plane. Calculations are performed within the local density
approximation (LDA)~\cite{Hohenberg}.  We use pseudopotential method
with mixed basis set\cite{Fu_mix}.  The Troullier-Martins type of pseudopotentials
are employed~\cite{Troullier}. Details for generating pseudopotentials,
including atomic configurations, pseudo/all-electron matching radii, and
accuracy checking, were described elsewhere\cite{Vpseudo}. The energy
cutoff is 100 Ryd, which is sufficient for convergence. The calculations
are performed in two steps: the optimized cell structure and atomic positions
are first determined by minimizing the total energy and Hellmann-Feynman
forces, and after the structural optimization, the polarization dispersion
of $\phi(\kpp)$ is calculated using the modern theory of polarization.\cite{King-Smith,Resta}
Our LDA-calculated inplane lattice constant for unstrained PT is $a$=3.88{\AA}, with
$c/a=1.04$, both agreeing well with other existing calculations.

\begin{figure}
\centering
 \includegraphics[width=10cm]{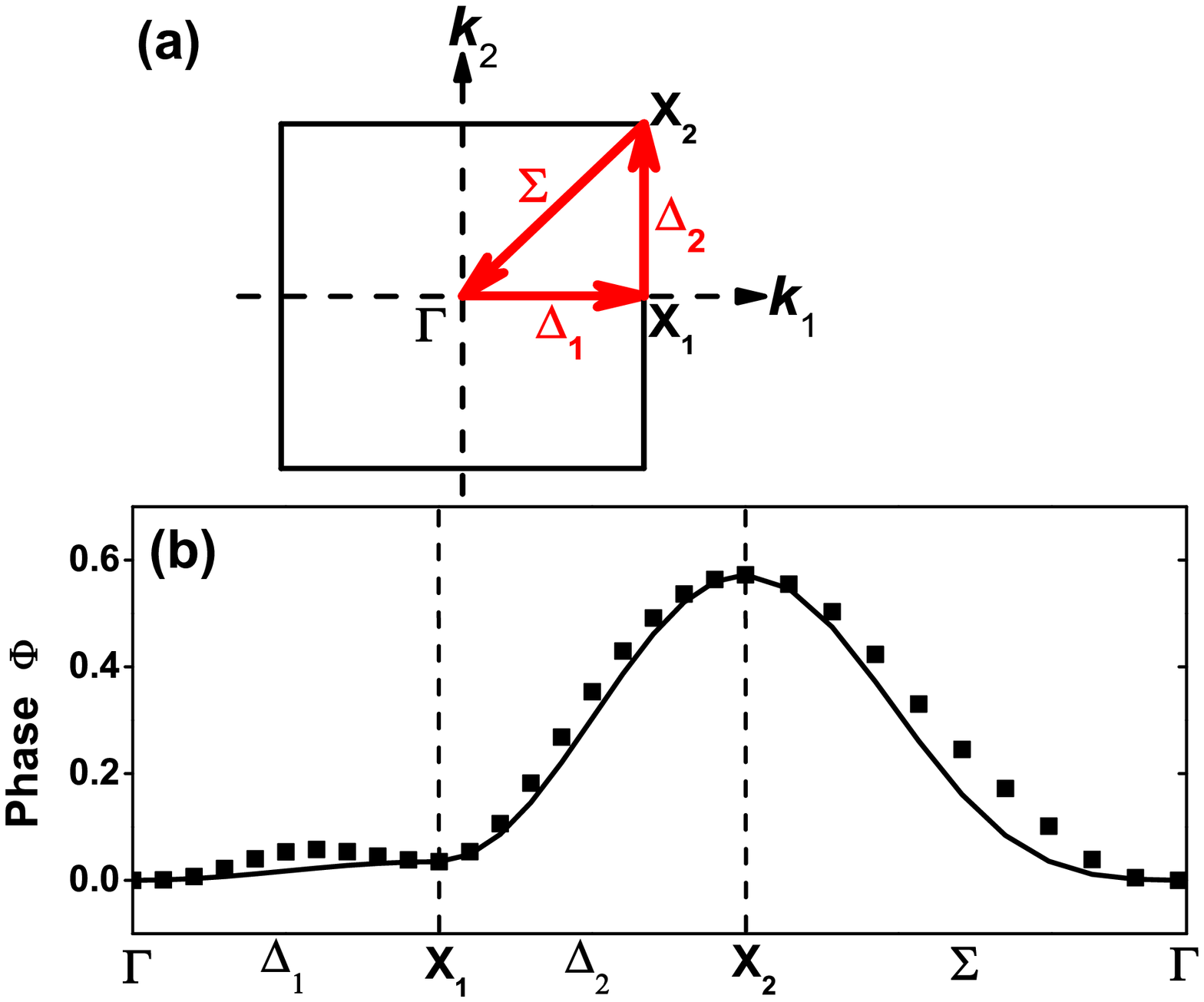}
 \caption{(a) The 2D Brillouin zone for the $\kpp $ plane; (b) Berry's phase
 at different $\kpp $ points for PbTiO$_3$ at equilibrium (symbols:
 direct calculation results; curve: analytical results). The $\phi(\kpp )$ phase
 is in units of radian.}
 \label{FPT}
\end{figure}

Figure \ref{FPT}(a) shows the reduced 2D Brillouin zone that the $\kpp$
points sample over. The calculated $\phi$ phases at individual $\kpp$ points along
the $\Gamma \rightarrow X_1 \rightarrow X_2 \rightarrow \Gamma$ path are given in Fig.\ref{FPT}(b).
Reciprocal-space coordinates of $X_1$ and $X_2$ are $\kpp=(\pi/a,0)$
and $(\pi/a, \pi/a)$, respectively. The dispersion curve is rigidly shifted such
that the phase at $\Gamma$ is taken as the zero reference.

Before we discuss the specific results in Fig.\ref{FPT}, we need to
point out that the shape of this $\kpp$-dependent phase curve is
translation invariant. As is known, the electronic polarization
alone can be an arbitrary value, if the solid is uniformly translated
with respect to a fixed origin of coordinates.
Though different translations will
change the absolute location of the polarization-dispersion curve,
the shape of the curve remains unaffected, however. This can be
easily illustrated by analyzing the change in the $\phi(\kpp)$ phase
when one displaces the solid arbitrarily. Let the
wave function of the original system be
$\psi_{n\vec{k}}(\ver)=e^{i\vec{k}\cdot\ver}u_{n\vec{k}}(\ver)$,
where $u_{n\vec{k}}(\ver)=u_{n\vec{k}}(\ver+\veR)$.  Now, we
displace the solid by an arbitrary vector $\vec{r_0}$ while the
origin of coordinates is fixed. Let us
denote the original system using script A and the displaced system
using script B, so $\vec{r}_B=\vec{r}_A+\vec{r}_0$. The wave
functions of the displaced system satisfy

\begin{equation}
\psi_{n\vec{k}}^B(\ver_B)=\psi_{n\vec{k}}^A(\ver_A)
=\psi_{n\vec{k}}^A(\ver_B-\ver_0) \ \ .
\end{equation}

\noindent
Thus we have $u_{n\vec{k}}^B(\ver_B)=
e^{-i\vec{k}\cdot\ver_0}u_{n\vec{k}}^A(\ver_B-\ver_0)$. Substituting this relation
into Eq.(\ref{EPhi_1}) or Eq.(\ref{EPhi_2}), one can obtain that
the $\phi(\kpp)$ of the displaced system is

\begin{equation}
 \phi^B(\vec{k}_\bot)=\phi^A(\vec{k}_\bot)+\ver_0\cdot\vec{G}_\| N_{band}^{occ} \ ,
\label{EPhi_BA}
\end{equation}
where $N_{band}^{occ}=M$ is the number of bands occupied by
electrons. The phase differences between the A and B systems are thus a
constant, independent of $\kpp$.

Several observations are ready in Fig.\ref{FPT}(b): (1) The largest
$\phi(\kpp)$ polarization does not come from the zone-center
$\Gamma$ point. Rather surprisingly, the largest
$\phi(\kpp)$ phase is from the $X_2$ point which lies at the far end
of the BZ. (2) The polarization curve is flat along the $\Gamma-X_1$
line, showing only a small dispersion. On the other hand, the
dispersion becomes very large along the $\Gamma-X_2$ direction. (3)
At $\kpp$ points of high symmetry (such as $\Gamma$, $X_1$, or
$X_2$), the curve in Fig.\ref{FPT}(b) has zero slope, similar to the
electron band structure. (4) The dispersion of polarization also
shows subtle details which could not be easily understood.  For
example, there is a local (though not very pronounced) maximum along
the $\Delta _1$ line, making the $X_1$ point a local minimum in both
$\Gamma-X_1$ and $X_1-X_2$ directions.

\begin{figure}[discontinuity]
\centering
 \includegraphics[width=10cm]{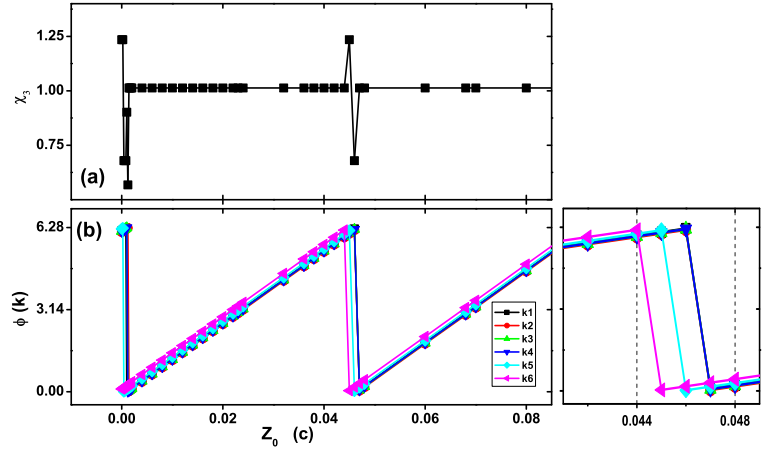}
 \caption{(Color online) (a) Total polarization in strained PbTiO$_3$ of inplane lattice
 constant $a=3.72${\AA} as a function of the uniform displacement $z_0$ of
 five atoms; (b) the $\phi(\kpp)$ phases at six Monhorst-Pack sampling
 $\kpp$ points as a function of $z_0$. For each $c/N^{occ}_{band}$ change
 in $z_0$, the $\phi(\kpp)$ phases change by $2\pi$. In (b), the $\phi(\kpp)$
 phase curves are enlarged in the right side of the figure for $z_0$ between
 0.044 and 0.048.}
 \label{FZo}
\end{figure}

Our calculations further reveal that, despite the fact that the
polarization in Fig.\ref{FPT}(b) exhibits substantial $\kpp$
dependency, the dispersion width ($\sim $0.6) is much smaller than
$2\pi$. This finding is important for the following reason. As
described in the introduction, if the differences of the
$\phi(\kpp)$ phases at different $\kpp$ points are greater than
$2\pi$, one would encounter a difficulty in determining which branch
of phase a specific $\kpp$ point should be assigned. This difficulty
can be avoided only after the phases of all $\kpp$ points are mapped
out. Fortunately, the result in Fig.\ref{FPT}(b) tells us that the
phase contributions from different $\kpp$ points are fairly close,
and the differences are far less than the critical value of $2\pi$
that may cause the above difficulty. Nevertheless, we should point
out that even a small polarization dispersion as in Fig.\ref{FPT}(b)
may still give rise to spurious results on total polarization. To
illustrate this, we displace all five atoms in PbTiO$_3$ along the
polar {\it c}-axis by a distance $z_0$. Fig.\ref{FZo}(a) shows the
total (electronic +ionic) polarization, computed from the geometric
phase, as a function of the displacement $z_0$ (in unit of $c$).
Intuition tells us that the total polarization should be uniquely
determined and translationally invariant. However, we see in
Fig.\ref{FZo}(a) that unphysical discontinuity happens for some
$z_0$ points, and this discontinuity shows up periodically. To
understand what causes the discontinuity, we examine the phase
contributions from individual $\kpp$ (sampled according to the Monkhorst-Pack
scheme\cite{Monkhorst}), depicted in Fig.\ref{FZo}(b). Figure
\ref{FZo}(b) shows that the individual-$\kpp$ phases indeed are a
periodic function of $z_0$, explaining why the discontinuity in
Fig.\ref{FZo}(a) is periodic. Here it may be useful to comment
briefly on the length of the periodicity. One might think that by
displacing the solid by a distance of $c$ in the
$c$-axis direction, the $\phi(\kpp)$ phase would change by a value
of $2\pi$. However, the periodicity in Fig.\ref{FZo} is much smaller
than $c$. The explanation is simple. As a matter of fact, in real space the individual
$\phi(\kpp)$ has a periodicity of $\frac{1}{N_{band}^{occ}}c$
(instead of $c$), which for PbTiO$_3$ the periodicity is 0.0455$c$
because $N_{band}^{occ}=22$. This is indeed consistent with the
numerical calculation in PT (Fig.\ref{FZo}b). The length of
periodicity can be seen from Eq.(\ref{EPhi_BA}), showing
that, whenever $\vec{r}_0=\frac{n}{N_{band}^{occ}}\vec{R}_\|$ ($n$
is an arbitrary integer and $\vec{R}_\|$ is the lattice vector along
the $\vec{G}_\|$ direction), the $\phi^B(\kpp)$ and $\phi^A(\kpp)$
differ by $\phi^B(\vec{k}_\bot)=\phi^A(\vec{k}_\bot)+2\pi n$.
Fig.\ref{FZo}(b) also reveals the reason responsible for the
discontinuity of the total polarization. Spurious discontinuity
occurs when the $\phi(\kpp)$ phases of some (but not all) individual
$\kpp$ exceed $2\pi$ [Fig.\ref{FZo}(b)]. Under this situation,
computers incorrectly shift the phases of these $\kpp$ points back
to the principle range, yielding spurious total polarization.
According to our experience, spurious polarization often takes place
in two circumstances: one is for materials of very large
polarization, such as tetragonal BiScO$_3$, and another is when
atoms in the unit cell are translationally shifted. Given the
small bandwidth of the $\phi(\kpp)$ dispersion, it is now straightforward
that, by using different $\vec{r}_0$s, we can avoid the spurious
polarization.  However for some materials, if the dispersion width
from different $\kpp$ points is larger than $2\pi$, one may have to
rely on the continuity of the $\phi(\kpp)$ phases, and map out the
phases of individual $\kpp$ points over the whole two-dimensional
$\kpp$ plane in order to find the correct phase branch.

\section{Strain dependence of polarization structure}

An important property of ferroelectrics is that the polarization is
strongly dependent on strain.  While strain can change the total
polarization, response of the polarization dispersion structure to
strain could also be an interesting problem. Here we investigate the
response of the polarization structure under inplane strain in
PbTiO$_3$. For each in-plane ($a$) lattice constant, the
out-of-plane $c$ lattice constant and atomic positions are fully
relaxed, by minimizing the DFT total energy. The polarization
structure is then determined using the optimal structure.

Figure \ref{FPT_str} shows the phase dispersion curves for PbTiO$_3$ at
different inplane lattice constants.  All curves are shifted so that
the phase at $\Gamma$ point is zero, in order to conduct direct
comparison. Three conclusions can be drawn from Fig.\ref{FPT_str}:
(1) The relative phase, $\phi (\kpp)-\phi(\Gamma )$, changes
drastically for $X_2$, but not so significantly for $X_1$. (2) At increasing strain,
(or smaller inplane $a$ constant), the bandwidth of the dispersion
initially changes very little when $a=3.84${\AA}, and then starts to
{\it decrease} upon further increasing strain to $a=3.80${\AA}. The
decline of the dispersion bandwidth is rather surprising, since a
compressive inplane strain is known to enhance the total
polarization in PT. The decline is also counterintuitive when one
considers that the decreasing inplane lattice constant makes the
atom-atom coupling stronger within the inplane directions, and
should therefore have increased the bandwidth. One possible reason
that may cause the decrease of the bandwidth is given in the next
section. As a result of the
declining dispersion, the polarization curve becomes notably
``flat'' at small $a=3.65${\AA}. (3) The curvature of the
dispersion also shows subtle changes, featured by the fact that a
new dispersion minimum appears along the $X_2-\Gamma$ line at large
strain. As a consequence, the dispersion curvature [i.e., the second
derivative $\bigtriangledown  ^2_{\kpp}P(\kpp)$] at $\Gamma $ point
alters its sign from being positive (at large $a$) to negative (at small $a$).
Furthermore, the local maximum between $\Gamma-X_1$ for unstrained
PT turns into a new minimum at large inplane strains. Meanwhile, the
$X_1$ point changes from a minimum into a saddle point, when strain
increases.

The calculations thus reveal that, while inplane strain has been
previously known to introduce interesting modifications (sometimes
markedly enlarged \cite{Ederer} and sometimes remarkably small
\cite{Lee}) to the {\it total} $c$-axis polarization, its effects on
the polarization dispersion at individual $\kpp$ points appear to be
even richer, showing that the polarization structure indeed worths
studying. The subtle response of the polarization structure, as
predicted above, indicate that there is new and rather complex
physics behind the results in Fig.\ref{FPT_str}. While we know that
the strain-induced changes in the polarization dispersion must be
associated with the fundamental modification of electron wave
functions, we also have to admit that the DFT results obtained in
our numerical calculations are puzzling, and an intuitive
understanding of the results is difficult for two reasons. First,
this is an early attempt to investigate the polarization structure,
and there is not much previous understanding in the literature.
Second, although Eq.(\ref{EPhi_1}) and Eq.(\ref{EPhi_2}) allow us to
compute precisely the polarization of individual $\kpp$, a direct
and more intuitive connection between $\phi(\kpp)$ and Bloch wave
functions is hard to capture from these equations. As a result, it
would be very helpful if one could find an alternative way to
understand the polarization structure and the computation results.
For instance, what determines the polarization at individual $\kpp$
point, and why $\phi(\kpp)$ maximizes at the $X_2$ point? In the
next section, we attempt a scheme which we wish to be able to offer
a more intuitive understanding of the polarization structure.

\section{Wannier function formulation of polarization structure}

As mentioned above, Eq.(\ref{EPhi_1}) and Eq.(\ref{EPhi_2}) give us
little intuitive sense on the direct $\kpp$ dependence of the Berry's phase. In
order to get more insight, we use Wannier functions to analyze the
polarization structure. Previously, Wannier functions have been
found very useful in analyzing real-space local
polarization\cite{Wu,Maxi}. Here we employ the Wannier-function
approach for a different purpose, namely to understand the
$\kpp$-dependence of the polarization structure. The Wannier functions
are defined as
\begin{equation}
W_n(\ver-\veR)=\frac{\sqrt{N}\Omega}{(2\pi)^3}
\int_{BZ}d\vec{k}e^{i\vec{k}\cdot(\ver-\veR)}u_{nk}(\ver)\ \,
\end{equation}
or
\begin{equation}
u_{nk}(\ver)=\frac{1}{\sqrt{N}}\sum_{\veR}
e^{-i\vec{k}\cdot(\ver-\veR)}W_n(\ver-\veR) \ \,
\label{EWa}
\end{equation}
where $\veR$ runs over the whole real-space lattice vectors.
By substituting Eq.(\ref{EWa}) into Eq.(\ref{EPhi_1}) and carrying
out analytically the integral over $\kpr$, it is
straightforward to derive, for tetragonal
perovskites, the polarization at individual $\kpp$ as
\begin{equation}
\phi(\kpp)=\frac{2\pi}{c}\sum_{\Rpp}\sum_{n=1}^M \int
\rpr W_n^*(\vec{r})W_n(\vec{r}-\Rpp)
e^{i\kpp \cdot \Rpp}d\vec{r} \ \,
\end{equation}
where $\rpr $ is the projection of vector $\ver$ along the
polarization direction, $\Rpp$ is the projection of lattice vector
$\veR$ onto the plane perpendicular to the polarization direction.
For convenience of discussion, we separate the sum over $\Rpp$ into
the $\Rpp=0$ term and the rests,
\begin{equation}
\phi(\kpp)=\phi_{0}+\frac{2\pi}{c}\sum_{\Rpp\neq0}\sum_{n=1}^M \int
\rpr W_n^*(\vec{r})W_n(\vec{r}-\Rpp)
e^{i \kpp \cdot \Rpp}d\vec{r}\ \,
\label{EWPhi}
\end{equation}
where for $\Rpp=0$,
$\phi_{0}=\sum_{n=1}^{M}\int(\vec{r})_{\parallel}
W_{n}^{*}(\vec{r})W_{n}(\vec{r})d\vec{r}$ is the phase contribution
from the same unit cell. Eq.(\ref{EWPhi}) is the basis for
understanding the polarization structure. From this equation, we
observe the following.

First, it is now clear that the $\kpp$-dependent part of $\phi(\kpp)$ comes
only from the $\Rpp \neq 0$ terms, which correspond to the overlap of the
Wannier functions in neighboring cells.  In other words, the $\kpp$
dependence of the $\phi(\kpp)$ phase results from the overlap of
the Wannier functions of different cells that are displaced by $\Rpp$
from each other within the plane that is perpendicular to the
direction of polarization. While the choice of the Wannier function is
known to be non-unique due to the gauge uncertainty, the sum of the
Wannier-function overlap over occupied bands is a uniquely defined
quality which does not depend on the gauge.  It is this quantity
that determines the shape of the polarization structure.

Second, Eq.(\ref{EWPhi}) explains why the bandwidth of polarization
dispersion is often much smaller than $2\pi$. Since only the second term
in this equation is $\kpp$ dependent, and since the Wannier
functions are generally well localized compared to the size of unit
cell, one expects the overlap $W^*_{n}(\ver)W_{n}(\vec{r}-\Rpp)$ to
be much smaller than unity for $\Rpp\neq0$. This is consistent with
our numerical results in Fig.\ref{FPT}, namely,
$\phi(\kpp )-\phi_{0} \approx 0.6 \ll 2\pi$.

Third, since the dispersion in $\phi(\kpp)$ comes from the overlap
of the Wannier functions between cells of different $\Rpp$s in the
{\it xy}-inplane directions, it explains why the polarization
structure is very sensitive to inplane strain, where by changing
inplane lattice constant, the distances between neighboring
cells are effectively altered. Meanwhile, we recognize that a
precise understanding of how the bandwidth depends on the inplane
strain is not as simple as one might think. Naively one tends to
think that, with the decline of inplane lattice constant, the
dispersion is to increase, since the overlap
$W_{n}^{*}(\ver)W_{n}(\vec{r}-\Rpp)$ increases when $\Rpp$
decreases. This will lead to the widening of the polarization
dispersion width, which is opposite to what we found in
Fig.\ref{FPT_str}. This puzzling contradiction can be resolved by
noticing that, in addition to being dependent on the overlap
strength between $W_{n}(\ver)$ and $W_{n}(\vec{r}-\Rpp)$ within the
{\it perpendicular} plane, the dispersion width also hinges on the
localization length ($l^{\rm WF}_{\parallel}$) of the Wannier
functions along the direction {\it parallel} to the polarization, as
a result of the $\rpr $ operator in Eq.(\ref{EWPhi}). With the
increasing inplane strain, the $l^{\rm WF}_{\parallel}$ is to
shrink. We thus see that the bandwidth of polarization is determined
by the balance of two competing factors between the increasing
Wannier-function overlap and the decreasing $l^{\rm WF}_{\parallel}$
localization length. When the latter dominates, the bandwidth
declines as we have seen in Fig.\ref{FPT_str} from numerical
calculations.

\section{Curve analysis}

With the general understanding of the polarization structure in the
above section, we next attempt to determine analytically the
polarization dispersion specifically for PbTiO$_3$, aimed to obtain
further insight into the  important details of the polarization
structure. As will become clear later, our analysis in the following
also explains what determines the $\phi(\kpp)$ polarization at
special points of $\Gamma $, $X_1$ and $X_2$.  We begin by defining
parameters
\begin{equation}
t(\Rpp)=\frac{2\pi}{c}\sum_{n=1}^M\int
\rpr W^*_n(\ver)W_n(\ver-\Rpp)d\ver \  ,
\label{EP_t}
\end{equation}
and then,
\begin{equation}
\phi(\kpp)=\sum_{\Rpp}t(\Rpp)e^{i\kpp \cdot \Rpp} \ .
\label{EP_TB}
\end{equation}

For dielectrics of insulating nature, Wannier functions are highly
localized, and decay exponentially with the distance
\cite{Kohn_WF,Marzari}. As a result, $t(\Rpp)$ also decay quickly
with the increase of $|\Rpp |$, so we can adopt the tight-binding like
approach and consider only several $\Rpp$s that correspond to some
nearest neighbors (NN). We consider up to the $2^{nd}$ NNs, where

\[\Rpp = \left\{ \begin{array}{ccc}
(0 &  0 )&  {\rm on\ site} \\
(\pm a & 0 )& {\rm 1NNs} \\
(0 & \pm a )& {\rm 1NNs} \\
(\pm a & \pm a )& {\rm 2NNs}  \\
\end{array}\right. \]
Taking advantage of tetragonal symmetry, we can rewrite
Eq.(\ref{EP_TB}) as
\begin{equation}
 \begin{split}
  \phi(\kpp)=t_0+
2t_1[\cos(k_1a) + \cos(k_2a)] \\
+ 2t_2[\cos(k_1+k_2)a + \cos(k_1-k_2)a ] \ ,
 \end{split}
\end{equation}
where $t_i$ is the $i^{th}$ NNs contribution defined in Eq.(10),
and $\kpp=(k_1,k_2)$. This expression gives us a more direct sense
of the $\phi(\kpp)\sim \kpp$ polarization dispersion, approximated to
the second nearest neighbors. At special $\kpp$ points of $\Gamma $,
$X_1$, and $X_2$, the phases are $\phi(\Gamma )=t_0+4t_1+4t_2$,
$\phi(X_1)=t_0-4t_2$, and $\phi(X_2)=t_0-4t_1+4t_2$, respectively.
We could thus clearly see that the $t_0$ term, corresponding to
$\Rpp=0$, acts to rigidly shift the polarization curve as a whole.
Meanwhile, the phase relative to the $\Gamma$ (i.e., the dispersion)
is determined by the $t_1$ and $t_2$ quantities, and more specifically,
\begin{eqnarray}
\phi(X_1)-\phi(\Gamma )=&-4t_1 & -8t_2 \ , \nonumber\\
\phi(X_2)-\phi(\Gamma )=&-8t_1 & \ .
\label{Edif}
\end{eqnarray}
These equations are useful, since they tell us that (1)
the relative height at $X_2$ (which contributes most to the
polarization in PT), $\phi(X_2)-\phi(\Gamma)$, is
determined by $t_1$, associated with the overlap of the Wannier
function in the $1^{st}$ NNs. $t_1<$0 for PbTiO$_3$ in equilibrium.
(2) Under the assumption that $t_2$ is negligible, $\phi(X_2)-\phi(\Gamma)$
will be larger than $\phi(X_1)-\phi(\Gamma)$ by a factor of 2.

Within the second nearest-neighbor approximation, one can further
determine analytically the dispersion along the $\Gamma \rightarrow
X_1 \rightarrow X_2 \rightarrow \Gamma $ line in the 2D Brillouin
zone as

\begin{widetext}
\[ \phi(\kpp)=\left \{
 \begin{array}{ll}
t_0+2t_1+(2t_1+4t_2)\cos(k_1a), & {\rm for }\ \Gamma \rightarrow X_1 \ {\rm with}\  k_2=0 \\
t_0-2t_1+(2t_1-4t_2)\cos(k_2a), & {\rm for }\ X_1 \rightarrow X_2 \   {\rm with }\  k_1=\pi /a \\
t_0+2t_2+4t_1\cos(k_1a)+2t_2\cos(2k_1a), & {\rm for }\ X_2
\rightarrow \Gamma \ {\rm with}\ k_1=k_2
 \ .
 \end{array}
\right. \]

\end{widetext} The polarization structure could thus be expressed as a simple
combination of cosine functions.

To examine whether the second-NN approximation is sufficient, we fit
the analytical results to the numerical DFT calculations to
determine the $t_i$ ($i=0,1,2$) parameters. Note that only
$\phi(\kpp)$s at three points (i.e., $\Gamma$, $X_1$ and $X_2$) are
fitted. The obtained $t_i$ values are given in Table \ref{Tpara}.
These values are then used to determine the whole dispersion curve,
shown in Fig.\ref{FPT}(b) for PbTiO$_3$ in equilibrium structure of
$a=3.88${\AA}. We could see that the analytical curve agrees well
with the DFT result, implying that the 2nd NN approximation works.
On the other hand, some fine structure of the curve (such as the
small local maximum along the $\Gamma-X_1$) can not be reproduced,
where for a better fitting, approximation beyond the 2nd NNs would
be necessary.

\begin{table}
\caption{The fitting $t_1$ and $t_2$ parameters for PbTiO$_3$ at different lattice
constants. $t_0$ is not shown here since it does not affect dispersion.}
\label{Tpara}
\begin{tabular}{ccc}
  \hline \hline
  a({\AA}) &  $t_1$ & $t_2$ \\
  \hline
  3.88  &-0.072 & 0.031 \\
  3.84  &-0.072 & 0.032 \\
  3.80  &-0.064 & 0.031 \\
  3.72  &-0.031 & 0.023 \\
  3.65  &-0.010 & 0.016 \\
  \hline\hline
\end{tabular}
\end{table}

From Table \ref{Tpara} one can also see how the $t_i$ quantities
are influenced by inplane strain. $t_1$ declines substantially as
$a$ decreases below 3.80{\AA}, while $t_2$ shows a less dependence
on inplane strain. This makes
sense since, by varying the inplane strain, the main effect lies
in altering the nearest-neighbor interaction among Wannier functions.
For $a>3.80${\AA}, $|t_1|$ approximately equals  2$|t_2| $, confirming
the importance of the nearest neighbor interaction. For large strains
of $a<3.72${\AA}, $|t_1|$ and $|t_2|$ become comparable, for which it
is likely that higher orders of NNs are also needed.

\section{Comparison with Barium Titanate}

It is of interest to compare the polarization dispersions between
BaTiO$_3$ (BT) and PbTiO$_3$ (PT), since these two substances have rather different tetragonality,
magnitude of polarization, and sizes of A-site atoms. For this
purpose, we have studied the polarization structure in BT, for which
a tetragonal symmetry is enforced so that a direct comparison with
PT can be made. Following the same procedure as for PT, we optimize
the cell structure and atomic positions of BT at different inplane
lattice constants, and calculate the corresponding polarization
structures.

\begin{figure}[BTstrain]
\centering
\includegraphics[width=12cm]{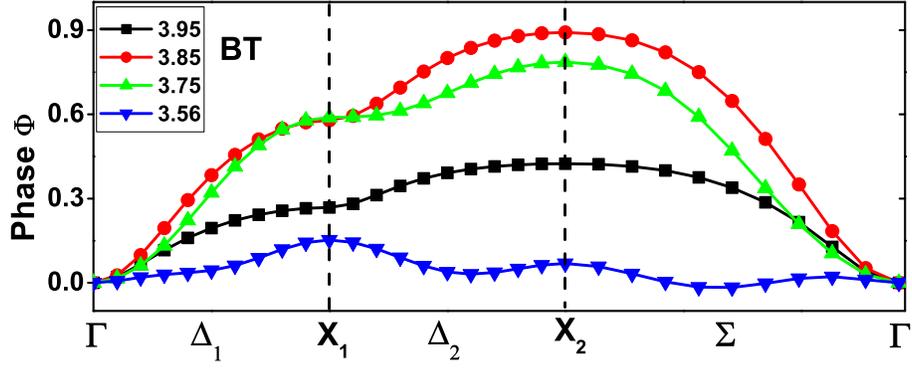}
 \caption{(Color online) Polarization dispersions for BaTiO$_3$ at different inplane lattice
 constants. Symbols are direct calculation results; lines are guide for eyes. }
\label{FBT}
\end{figure}

Fig.\ref{FBT} displays the polarization structure for BaTiO$_3$ at
different inplane lattice constants. Let us first focus on the
dispersion of the equilibrium BaTiO$_3$. The LDA-calculated
equilibrium inplane lattice constant of BT is $a=3.95${\AA}.  Apart
from similarities to PT (e.g., $\phi $ maximizes at $X_2$), our
calculations reveal some interesting differences between PT and BT
under zero strain: (1) The BT dispersion curve has a significantly
smaller bandwidth ($\sim $0.42) than that of PT ($\sim $0.57). Since
the bandwidth is determined by the difference
$\phi(X_2)-\phi(\Gamma)$, i.e., by $t_1$, a smaller bandwidth
indicates less overlapping Wannier's functions between nearest
neighbors in BaTiO$_3$, which could be explained by the larger
inplane lattice constant $a$ for BT at equilibrium. (2) Unlike PT,
the polarization in BT is not small at $X_1$. This again can be
attributed to the large inplane lattice constant in BT, which leads
to a negligible contribution from the 2nd NNs, i.e., $t_2$ is small
in BT. Indeed, we numerically found that $t_2$ is -0.007 in BT,
compared to 0.031 in PT. By Eq.(\ref{Edif}), $\phi(X_1)$ is about
half of the $\phi(X_2)$ value if $t_2$ is small, which is indeed
born out in Fig.\ref{FBT}. (3)As a consequence of observation (2),
the dispersions of BT and PT along the $\Gamma \rightarrow X_1$ are
not quite similar. There is a local maximum between $\Gamma-X_1$ for
PT, whereas for BT, no local maximum exists and $X_1$ becomes a
saddle point.

\begin{figure}[x1x2fit]
\centering
 \includegraphics[width=10cm]{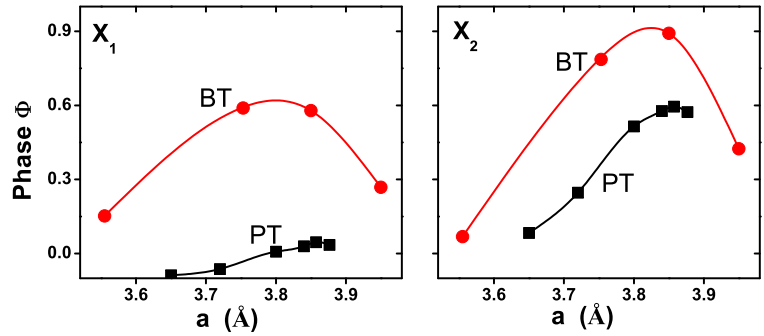}
 \caption{Dependencies of the $\phi(\kpp)$ phases at $X_1$ point (left) and at $X_2$ point
 (right) as a function of inplane lattice constant, for PT and BT.}
 \label{Fx1x2}
\end{figure}

Upon strain, BaTiO$_3$ and PbTiO$_3$ exhibit sharp difference in
their strain dependence of dispersion bandwidth. As we saw
previously in Fig.\ref{FPT_str}, inplane strain causes the bandwidth
declining for PbTiO$_3$.  However, for BaTiO$_3$, a dramatic {\it
enlargement} in bandwidth occurs, when $a$ decreases from 3.95{\AA}
to 3.85{\AA}. The bandwidth maintains a large value at
$a$=3.75{\AA}, after which it starts to drop. In BaTiO$_3$ the
polarization dispersion bandwidth thus shows an interesting
non-monotonous dependence on inplane strain. This characteristic
non-monotonous dependence strongly supports our conjecture that the
two competing factors determine the bandwidth, as described above
in Section IV.
When strain is small in BT, the overlapping of Wannier functions
located at the nearest neighboring $\Rpp$s plays a dominant role,
and the increasing overlap leads to a larger $|t_1|$ and thus larger
bandwidth. As inplane strain becomes large ($a<3.85$\AA), the
atom-atom interaction along the $c$-axis is considerably weakened
due to elongated $c$-lattice length.  As a consequence, the
shrinking $l^{\rm WF}_{\parallel}$ localization length of Wannier
functions along the $\rpr$ direction takes over and becomes
dominant, giving rise to the declining bandwidth.  This, once again,
reveals that the polarization dispersion contains rich information.
To make more quantitative comparison, we replot in Fig.\ref{Fx1x2}
the strain dependence of the $\phi(\kpp)$ phases at $X_1$ and $X_2$,
relative to the $\Gamma$ point.
Fig.\ref{Fx1x2} is of some useful value since it allows us to contrast the
$\kpp$-specific polarizations in two materials at {\it the same}
fixed inplane lattice constant. The difference between BT and PT is
thus not related in a significant sense to atom-atom distance, but
largely due to the overlap of respective Wannier functions. In
Fig.\ref{Fx1x2},  both $\phi(X_1)$ and $\phi(X_2)$ are seen to be
far greater in BaTiO$_3$ than in PbTiO$_3$, for a fixed $a$
constant. The greater values of $\phi(\kpp)$ in BT could possibly
originate from the fact that the Wannier functions in this material
is more spreading due to the larger size of Ba atom.

From the comparison between PT and BT, we could see that the
polarization structure has some common features for materials with
similar structure, and meanwhile, some distinctions revealing the
identities of materials. The common features allow us to understand
the polarization structure in general, just as for band structure,
most III-V semiconductors have direct band gaps. Differences in
polarization structure manifest the electron wave functions and
interatomic interactions on microscopic scale.

\section{Summary}

Two different approaches are employed to study the polarization
structure in perovskite ferroelectrics. Numerically we use the
density functional total-energy calculations and the modern theory
of polarization. Analytically we formulate a scheme to describe the $\kpp$
dependence of the polarization phase using Wannier functions. By
parameterizing the Wannier-function overlapping, we further identify
the quantities that determine the $\phi(\kpp)$ phases at special
$\kpp$ points of interest. Our specific findings are summarized in
the following.

For PbTiO$_3$ at equilibrium, (i) the $\phi(\kpp)$ phase maximizes
at the Brillouin zone boundary of the 2D $\kpp$ plane, not the zone
center. (ii) The polarization structure shows little dispersion
along the $\Gamma-X_1$ line. However, the dispersion is large along
the $\Gamma-X_2$.  (iii) The bandwidth of the dispersion curve is
far below 2$\pi$. The small dispersion considerably eases the
difficulty in assigning the correct branch of individual $\kpp$
phase, but caution still needs to be taken when the $\phi(\kpp)$
phase is approaching $2\pi$.

Analytically, (iv) the expression, Eq.(\ref{EWPhi}), is given as the basis for understanding
the polarization structure. It also explains why the polarization
bandwidth is small compared to $2\pi$. (v) The polarization phase at individual
$\kpp$ is revealed to depend on the competition of two factors, namely the
overlapping strength of Wannier functions within the perpendicular $\veR_{\perp}$
plane and the localization length $l^{\rm WF}_{\parallel}$ of these Wannier functions.
(vi) Within the 2NN approximation, the $\phi(X_1)$ and $\phi(X_2)$ values in
ferroelectric perovskite are found to be
$ \phi(X_1)-\phi(\Gamma )=-4t_1 - 8t_2$, $\phi(X_2)-\phi(\Gamma )=-8t_1$.
If $t_2$ is negligible, the latter is 2 times of the former.
(vii) When PbTiO$_3$ is under compressive inplane strain, the polarization
bandwidth is found to decrease, whereas the total polarization increases.
The declining bandwidth implies that the localization length $l^{\rm WF}_{\parallel}$
of Wannier functions plays a dominating role in PbTiO$_3$.

By comparing BaTiO$_3$ with PbTiO$_3$, we show (viii) the equilibrium BT exhibits a
smaller bandwidth of 0.42, as compared to the bandwidth of 0.57 in PT.
(ix) $\phi(X_1)$ in BaTiO$_3$ is not small, unlike PT. The difference comes from
the fact that $t_2$ is negligible in BT, leading to the result that $\phi(X_1)$
is about half of the value of $\phi(X_2)$. But in PT, $t_2$ can not be neglected,
and acts to offset the $t_1$ contribution, giving rise to smaller $\phi(X_1)$ and
flat dispersion along the $\Gamma-X_1$ line. (x) As BaTiO$_3$ is under increasing
inplane strains, its polarization bandwidth displays a characteristic non-monotonous
variation by first increasing dramatically and then declining. The finding
lends a support to the qualitative understanding that two competing factors determine the
$\phi(\kpp)$ phase. (xi) When BaTiO$_3$ and PbTiO$_3$ are constrained to the
same inplane lattice constant, the $\phi(X_1)$ and $\phi(X_2)$ are shown to be
significantly larger in BT than in PT, unlike the case when two materials are in
equilibrium.

\begin{figure}[PTstrain]
\centering
 \includegraphics[width=12cm]{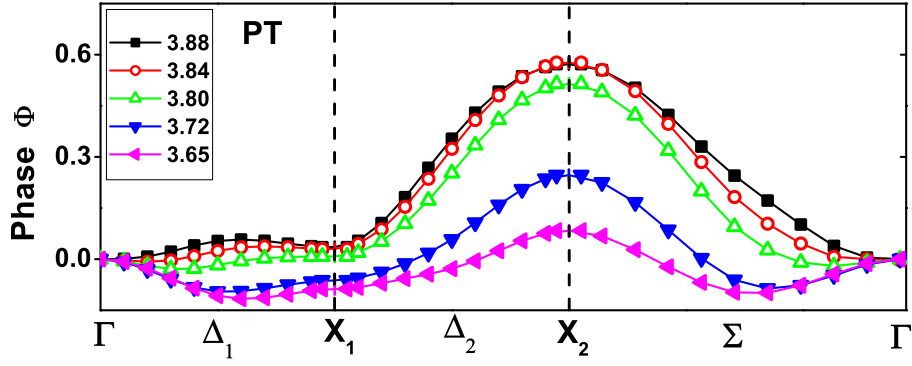}
 \caption{(Color online) The $\phi $ phases of different $\kpp$-points, for PbTiO$_3$
 under different inplane lattice constants. Symbols are direct calculation
 results; curves are guides for eyes. }
 \label{FPT_str}
\end{figure}

We conclude by pointing out that there are still many aspects of polarization structure
we do not yet understand. For example, we have not pursued beyond the 2nd
nearest neighbors to explain the local maximum between $\Gamma $
and $X_1$ in unstrained PT. We also do not know the physical significance
when $\phi(X_1)$ changes from a local minimum to a saddle point as displayed
in Fig.\ref{FPT_str} for PbTiO$_3$ under strains. We believe that further analysis of the
polarization structure could yield better knowledge on the
physics of dielectrics. Like band structure of solids, we hope that the polarization
structure can provide us a new tool of studying ferroelectric materials and
properties.

This work was supported by the Office of Naval Research.

\end{document}